\documentclass[12pt]{article}
\usepackage{latexsym}
\def\bseq{\begin{subequation}}  
\def\eseq{\end{subequation}}
\def\bsea{\begin{subeqnarray}}  
\def\esea{\end{subeqnarray}}

\newcommand{\beq}{\begin{equation}}
\newcommand{\eeq}{\end{equation}}
\newcommand{\bea}{\begin{eqnarray}}
\newcommand{\eea}{\end{eqnarray}}
\newcommand{\bdm}{\begin{displaymath}}
\newcommand{\edm}{\end{displaymath}}
\newcommand{\ba}{\begin{array}}
\newcommand{\ea}{\end{array}}
\newcommand{\ben}{\begin{enumerate}}
\newcommand{\een}{\end{enumerate}}
\newcommand{\bde}{\begin{description}}
\newcommand{\ede}{\end{description}}
\newcommand{\nn}{\nonumber}

\renewcommand{\r}{\right}
\renewcommand{\l}{\left}

\newcommand{\complesso}{{\ \hbox{{\rm I}\kern-.6em\hbox{\bf C}}}}
\newcommand{\reale}{{\hbox{{\rm I}\kern-.2em\hbox{\rm R}}}}
\newcommand{\1}{ \,  \raisebox{+0.14em}{{\hbox{{\rm \scriptsize ]}} \raisebox{-0.2em}{\kern-.8em\hbox{1}}}} \, }  

\newcommand{\p}{\partial}

\renewcommand{\a}{\alpha}
\renewcommand{\b}{\beta}
\newcommand{\g}{\gamma}

\renewcommand{\d}{\delta}

\newcommand{\e}{\epsilon}

\renewcommand{\l}{\lambda}

\newcommand{\m}{\mu}

\newcommand{\n}{\nu}

\renewcommand{\r}{\rho}

\newcommand{\s}{\sigma}

\renewcommand{\t}{\theta}

\begin{document}
\begin{titlepage}
\begin{flushright}
IFUM-740-FT\\
December 2002
\end{flushright}
\vspace{2cm}
 
\noindent{\Large \bf Noncommutative deformation of four dimensional }\\
\vspace{-0.3 cm}
\\
{\Large \bf Einstein  gravity}

\vspace{1cm}
{\bf \hrule width 16.cm}
\vspace {1cm}

\noindent{\large \bf Matteo A. Cardella and Daniela Zanon}

\vskip 2mm

{ \small \noindent Dipartimento di Fisica dell'Universit\`a di Milano and

\noindent INFN, Sezione di Milano, Via Celoria 16, 20133 Milano, Italy}
\vfill
\begin{center}
{\bf Abstract}
\end{center}
{\small We construct a model for noncommutative gravity in four dimensions, which reduces to the 
Einstein-Hilbert action in the commutative limit. Our proposal is based on a gauge formulation of gravity 
with constraints. While the action is metric independent, the constraints insure that it is not 
topological. We find that the choice of the gauge group and of the constraints are crucial to recover a 
correct deformation of standard gravity. 
Using the Seiberg-Witten map the whole theory is described  in terms of the vierbeins and of the Lorentz transformations   of 
its commutative counterpart. 
We solve explicitly the constraints and exhibit the first order noncommutative corrections to the Einstein-Hilbert action.} 
\vspace{2mm} 
\vfill \hrule width 6.cm
\begin{flushleft}
e-mail: matteo.cardella@mi.infn.it\\
e-mail: daniela.zanon@mi.infn.it
\end{flushleft}
\end{titlepage}

Up to now a consistent formulation of four dimensional noncommutative gravity that reduces 
to the standard Einstein-Hilbert theory in the commutative limit has proven quite difficult to approach \cite{Garcia-Compean:2002ss}, \cite{noncommactions}, \cite{deforming}, \cite{complex} \cite{Moffat:2000fv}, \cite{Vacaru:2000yk}. 
Among others there are problems like finding an invariant measure, solving the inconsistencies of a 
complex metric, singling out the correct degrees of freedom. In two and three dimensions these 
difficulties can be avoided since a theory of gravity can be formulated as a gauge theory and we know how 
to deform gauge transformations in a noncommutative geometry \cite{noncommin3d}, \cite{Cacciatori:2002cr}, \cite{Banados:2001xw}, \cite{noncommin2d}.

In this paper we pursue the idea of searching for a description of four dimensional noncommutative gravity as a gauge 
invariant theory with constraints. For the standard Einstein action with a 
cosmological term \cite{azione1}, \cite{azione2}  this can be easily done using the $SO(1,4)$ de Sitter group as a start, 
and then reducing the symmetry to the 
$SO(1,3)$ Lorentz group  via the torsion free constraint. The constraints play an important role: they 
allow to write an action which, although independent of the metric, is not topological \cite{vincoli}. Moreover their 
solution eliminates the unphysical, dependent degrees of freedom leaving the metric as the only dynamical 
field. 

When we consider  a noncommutative deformation of the theory introducing the Moyal $\star$-product \cite{starprod}, 
we have to face the fact that the  only  consistent gauge groups are the unitary groups. Thus we look for 
the simplest unitary groups which contain $SO(1,4)$ and $SO(1,3)$ with the aim to deform their algebra 
with the $\star$-product. The appropriate groups turn out to be  $U_{\star}(2,2)$ and 
$U_{\star}(1,1)\times U_{\star}(1,1)$ respectively  \cite{noncommactions}. In fact we find that starting from a gauge theory 
invariant under $U_{\star}(2,2)$ we can impose constraints that reduce the symmetry to a subclass 
contained in $U_{\star}(1,1)\times U_{\star}(1,1)$. We name this subalgebra $SO_{\star}(1,3)$ 
since it represents 
\emph{the simplest noncommutative deformation} of the Lorentz algebra $SO(1,3)$. We use the constraints to 
express the dependent gauge fields in terms of the independent ones and construct an action invariant under 
$SO_{\star}(1,3)$. It reduces to the standard action in the commutative limit.
Then we show that via the Seiberg-Witten map \cite{SW}, gauge transformations $\l$ in $SO(1,3)$ turn 
precisely into $SO_{\star}(1,3)$ transformations $\hat{\l}$. We define our noncommutative theory based on 
this set $\hat{\l}$ of gauge transformations. Thus we are allowed to express the $\theta$-dependence of the
fields in the action in terms of the fields in the commutative theory 
using the Seiberg-Witten map. In this fashion the whole theory is described in terms of the vierbeins 
and the Lorentz transformations.
 Finally we solve explicitly the constraints to first order in $\theta$ and exhibit the first 
order noncommutative correction to the Einstein-Hilbert action. 

\vspace{.8cm} We start by studying how the introduction of the $\star$ product leads to a  
deformation of the Lorentz group $SO(1,3)$ that we call $SO_{\star}(1,3)$.

 In a noncommutative theory, under an infinitesimal gauge transformation $\tilde{\l}$ 
 the  gauge connection $\tilde{A}_{\m}$ transforms  as follows: 
 \beq \hat{\d}_{\tilde{\l}}\tilde{A}_{\m} =  \p_{\m}\tilde{\l} + [\tilde{\l} , 
 \tilde{A}_{\m}]_{\star} = \p_{\m}\tilde{\l} + 
 \tilde{\l} \star \tilde{A}_{\m} -  \tilde{A}_{\m}\star \tilde{\l} 
 \label{transfconn}
 \eeq
 Since
 \beq [\hat{\d}_{\tilde{\l}_{1}} , 
 \hat{\d}_{\tilde{\l}_{2}}]\tilde{A}_{\m} = \hat{\d}_{[\tilde{\l}_1 , \tilde{\l}_2 
 ]_{\star}}\tilde{A}_{\m} \eeq
in order to have a representation of the Lie algebra of the group $[\l_1 , \l_2]_{\star}$ 
must be in the algebra.

Under the $\star$-operation the Lorentz algebra does not close. To prove this 
 we consider the basis of  a Clifford algebra 
 $\{ \gamma_{a}, \gamma_{b} \} = 2\eta_{ab}$, with  $\gamma_a$ $4\times 4$ matrices  and 
  $\eta_{ab}$ the flat Minkowski metric.
Then we define  $\gamma_{ab} \equiv \frac{1}{2}(\gamma_{a}\gamma_{b} - \gamma_{b}\gamma_{a})$ as the six 
generators of the Lorentz group, $\gamma_{5} \equiv i\gamma_{0} \gamma_{1} \gamma_{2} \gamma_{3}$, and construct
the Moyal commutator of two local Lorentz transformations $\l_1=\l_{1}^{ab}\gamma_{ab}$,  
$\l_2=\l_{2}^{ab}\gamma_{ab}$
 \bea [\l_1 , \l_2 ]_{\star} &=& \l^{ab}_{1} \star 
\l_{2}^{cd}\gamma_{ab}\gamma_{cd} -  \l_{2}^{cd} \star \l_{1}^{ab}\gamma_{cd}\gamma_{ab}\nn \\ & 
=&  \l_{1}^{ab} \star_{s}\l_{2}^{cd}[\gamma_{ab} , \gamma_{cd}] +  \l_{1}^{ab} 
\star_{a}\l_{2}^{cd}\{\gamma_{ab} , \gamma_{cd}\}   \label{1} 
\eea 
where \cite{noncommactions}  
\bea
f\star_{s} g & \equiv & \frac{1}{2}( f\star g + g \star f) = f g + 
\frac{1}{2!}\bigg{(}\frac{i}{2}\bigg{)}^{2}\t^{\a\b}\t^{\g\d}\p_{\a}\p_{\g}f\p_{\b}\p_{\d}g \nn \\
&+&~{\rm even ~powers ~in ~\theta} \nn \\
f\star_{a} g & \equiv & \frac{1}{2}( f\star g - g \star f) = 
\bigg{(}\frac{i}{2}\bigg{)}\t^{\a\b}\p_{\a}f\p_{\b}g + \nn \\ &+& 
\frac{1}{3!}\bigg{(}\frac{i}{2}\bigg{)}^{3}\t^{\a\b}\t^{\g\d}\t^{\r\s}\p_{\a}\p_{\g}\p_{\r}
f\p_{\b}\p_{\d}\p_{\s}g 
+  ~{\rm odd ~powers~ in~ \theta}
\eea
Since
 \bea
\lbrack  \gamma_{ab} ,  \gamma_{cd}  \rbrack  & =&  2( \eta_{ad}\gamma_{bc}  +  \eta_{cb}\gamma_{ad} 
-  \eta_{ac}\gamma_{bd} -  \eta_{bd}\gamma_{ac}) \nn \\
\{ \gamma_{ab} ,  \gamma_{cd} \} &=& -2i(\eta_{ac}\eta_{bd}i\1 + \e_{abcd}\gamma_{5}) \eea 
then
\bea 
[\l_1 , \l_2 ]_{\star} &=&(\l^{ab} + \l^{ab(2)} +...+ \l^{ab(2n)} +...)\gamma_{ab}+ \label{2}\\
&+& (\l^{1(1)}  +   
\l^{1(3)}+ ...+ \l^{1(2n +1)} + ...)i\1  + (\l^{5(1)} +\l^{5(3)} + 
...+\l^{5(2n +1)} +...)\gamma_{5}  \nn 
\eea
where  $\l^{(n)}$ is of order    $\t^{n}$.

The gauge transformations in (\ref{2}) are not Lorentz transformations but they have a special
and rather simple 
$\theta$-expansion: the terms which are
even powers in $\theta$ are Lorentz transformations, while the ones odd in $\theta$  have 
non vanishing components on $i\1$ and $\gamma_{5}$.
The set in (\ref{2}) forms a  subclass of the $U_{\star}(1,1) \times U_{\star}(1,1)$ 
algebra which is closed under the $\star$-product:   indeed it is easy to prove that if  
$\tilde{\l}_{1}$ and  $\tilde{\l}_{2}$  have a $\theta$-expansion as the one in (\ref{2}), then their 
Moyal-commutator  $ [\tilde{\l}_1 , \tilde{\l}_2 ]_{\star}$ has again the same kind of $\theta$-expansion, 
i.e. 
even powers are proportional to $\gamma_{ab}$, odd powers are proportional to $i\1$ and $\gamma_{5}$.
We call this subalgebra $SO_{\star}(1,3)$: it should describe the invariance of our 
noncommutative theory of gravity. 

In order to achieve this goal we start with a
$U_{\star}(2,2)$ gauge theory and 
break the symmetry to $SO_{\star}(1,3)$ imposing suitable constraints. The procedure is most easily 
elucidated for the commutative theory \cite{noncommactions}: in this case one  considers
  a $U(2,2)$ gauge theory and  breaks the symmetry down to $SO(1,3)$, obtaining a gauge formulation for standard gravity.
$U(2,2)$  is the Lie  group of complex $4 \times 4$ matrices  $U$ such that: $ U^{\dag}\tilde{\eta}U = \tilde{\eta}$ with
  $\tilde{\eta} = diag(+ + - -)$.
A basis of the  Lie algebra is given by 16 linear indipendent matrices $\l$ satisfying the relation:
\beq
 \l^{\dag} = - \tilde{\eta} \l^{\dag} \tilde{\eta}. \label{Lie}
\eeq  

In the Dirac-Pauli representation with $\gamma_{0}=\tilde{\eta}$ we choose the following basis $\tau^I$:

 \beq
(i\1 , \gamma_{5} , i\gamma_{a}^{+} , i\gamma_{a}^{-} , \gamma_{ab} ) \label{base}
\eeq 
where in addition to the generators of $U(1,1)\times U(1,1)$ one has
$\gamma_{a}^{\pm} \equiv\gamma_{a}(1 \pm \gamma_{5})$.\\

The connection $A_{\m}$ of the corresponding
gauge theory   is Lie algebra valued
 \beq 
 A_{\m} = a_{\m}i\1 + b_{\m} \gamma_{5} +
e^{a+}_{\m}i\gamma_{a}^{+} +  e^{a -}_{\m}i\gamma_{a}^{-} + \frac{1}{4}\omega^{ab}_{\m}
\gamma_{ab}
 \eeq
The field strength is
\beq
 F_{\m\n} \equiv \p_{\m}A_{\n} + A_{\m} A_{\n} - \m  \rightarrow \n 
=   F_{\m\n}^I \tau_I
\eeq
with components 
 \bea
F_{\m\n}^{1}& =& \p_{\m}a_\n   -  \m \leftrightarrow \n \nn \\
F_{\m\n}^{5} & =&  \p_{\m}b_\n    + \frac{1}{2}( e^{a+}_{\m} e^{-}_{\n a} 
-  e^{a-}_{\m} e^{+}_{\n a})  -  \m \leftrightarrow \n          \nn \\
F^{a+}_{\m\n}  & =& \p_{\m}e^{a +}_{\n}  - 2 b_{\m}e^{a+}_{\n}  + \omega^{ab}_{\m} e^{+}_{\n b}    
 -  \m \leftrightarrow \n   \nn \\
F^{a-}_{\m\n}   & =&  \p_{\m}e^{a-}_{\n}  +  2 b_{\m} e^{a-}_{\n} + \omega^{ab}_{\m} e^{-}_{\n b}   
-  \m \leftrightarrow \n    \nn \\
F ^{ab}_{\m\n} &=& \p_{\m}\omega^{ab}_{\n} +  \omega^{a}_{\m c} \omega^{cb}_{\n}
-4(e^{a+}_{\m}e^{b-}_{\n} + e^{a-}_{\m}e^{b+}_{\n})  -  \m \leftrightarrow \n 
\label{curva}
\eea
One can show that imposing the constraints
 \bea
 &&a_{\m} =b_{\m}=0 \nn \\
&&e^{a -}_{\m} = \b e^{a +}_{\m} \nn \\
&&F^{a+}_{\m\n} = 0 
\label{vinccomm} 
\eea
the gauge group $U(2,2)$ is broken into $SO(1,3)$ with an additional U(1) global symmetry.

Now we consider the following $SO(1,3)$ invariant action
 \beq 
 S = \int d^{4}x ~\e^{\m\n\r\s} Tr(\gamma_{5} F_{\m\n} F_{\r\s})
\label{action} 
\eeq
Using the constraints (\ref{vinccomm})  and the definition
\beq
R ^{ab}_{\m\n} = \p_{\m}\omega^{ab}_{\n} +  \omega^{a}_{\m c} \omega^{cb}_{\n}  -  \m \leftrightarrow \n
\label{Rordinezero} 
\eeq 
the action (\ref{action}) becomes
 \beq
S= - \frac{i}{2} \int d^{4}x ~\e^{\m\n\r\s}\e_{abcd}(R^{ab}_{\m\n} 
 -  8\b e^{a+}_{\m}e^{b+}_{\n})(R^{cd}_{\r\s} -  8\b e^{c+}_{\r}e^{d+}_{\s})
\eeq
The case $\b = 0$ gives the topological Gauss-Bonnet term , 
 while  $\b \ne  0$ gives also the  classical  Einstein action with a cosmological term.\\
It is worth noticing  that the choice  of the constant $\b$ in  (\ref{vinccomm}) determines the value of the cosmological constant of the model. Once the constraints are imposed  we are left with  $P_{a}\equiv i\gamma_{a}^{+} + \b i\gamma_{a}^{-}$ and  $M_{ab}\equiv \gamma_{ab}$,  a basis for the de Sitter or anti de Sitter group
depending on the sign of $\b$.  For $\b = 0$ one would obtain  the Poincar\`e group, but 
in this  case the action  (\ref{action}) becomes topological and it cannot be used to describe 
four dimensional gravity.

\vspace{0.8cm}

Now we turn to the noncommutative case. We consider a $U_{\star}(2,2)$  noncommutative gauge theory 
(i.e. a $U(2,2)$-gauge theory 
in a noncommutative space) and  we impose constraints to reduce the symmetry to 
$SO_{\star}(1,3)$. This we want to be the gauge symmetry of our noncommutative 
gravity, since, as emphasized above,
$SO_{\star}(1,3)$ is the {\em natural} noncommutative deformation of the ordinary 
 $SO(1,3)$ Lorentz algebra. 
 
In the  $U_{\star}(2,2)$ gauge theory  we write the connection  $\tilde{A}_{\m}$  as
 \beq 
 \tilde{A}_{\m} = \tilde{a}_{\m}i\1 + \tilde{b}_{\m} \gamma_{5} +\tilde{e}^{a
+}_{\m}i\gamma_{a}^{+} +\tilde{e}^{a-}_{\m}i\gamma_{a}^{-} +
\frac{1}{4}\tilde{\omega}^{ab}_{\m} \gamma_{ab} 
\eeq
where all the fields are functions of the space time coordinates and of the noncommutative parameter
$\theta$.
The corresponding field strength is given by
\beq
 \tilde{F}_{\m\n} = \p_{\m}\tilde{A}_{\n} + \tilde{A}_{\m}\star \tilde{A}_{\n} - \m  \rightarrow \n 
 =   \tilde{F}_{\m\n}^{I}\tau_I
 \eeq 
with components 
  \bea
\tilde{F}_{\m\n}^{1}& =& \p_{\m}\tilde{a}_\n   + i\tilde{a}_\m \star_{a} \tilde{a}_\n 
-  i\tilde{b}_\m \star_{a} \tilde{b}_\n  + \frac{i}{2}(\tilde{e}^{a+}_{\m} \star_{a} \tilde{e}^{-}_{a\n} 
+  \tilde{e}^{a-}_{\m} \star_{a} \tilde{e}^{+}_{a\n}) 
- \frac{i}{8}\tilde{\omega}^{ab}_{\m} \star_{a}\tilde{\omega}_{\n ab} -  \m \leftrightarrow \n \nn \\
\tilde{F}_{\m\n}^{5} & =&  \p_{\m}\tilde{b}_\n + 2i \tilde{a}_{\m}   \star_{a}\tilde{b}_{\n}   
+ \frac{1}{2}( \tilde{e}^{a+}_{\m} \star_{s}\tilde{e}^{-}_{\n a} 
-  \tilde{e}^{a-}_{\m} \star_{s}\tilde{e}^{+}_{\n a})   
 - \frac{i}{8} \e_{abcd} \ \tilde{\omega}^{ab}_{\m} \star_{a}\tilde {\omega}^{cd}_{\n} 
 -  \m \leftrightarrow \n          \nn \\
\tilde{F}^{a+}_{\m\n}  & =& \p_{\m}\tilde{e}^{a +}_{\n}  - 2\tilde{b}_{\m} \star_{s}\tilde{e}^{a+}_{\n}  
+ \tilde{\omega}^{ab}_{\m} \star_{s}\tilde{e}^{+}_{\n b}  + 2i\tilde{a}_{\m} \star_{a}\tilde{e}^{a+}_{\n} 
 + \frac{i}{2}\e^{a}_{\ bcd}\tilde{e}^{b+}_{\m} \star_{a}  \tilde{\omega}^{cd}_{\n}  
 -  \m \leftrightarrow \n   \nn \\
\tilde{F}^{a-}_{\m\n}   & =&  \p_{\m}\tilde{e}^{a-}_{\n}  +  2\tilde{b}_{\m} \star_{s}\tilde{e}^{a-}_{\n} 
+ \tilde{\omega}^{ab}_{\m} \star_{s}\tilde{e}^{-}_{\n b} + 2i\tilde{a}_{\m} \star_{a}\tilde{e}^{a-}_{\n} 
- \frac{i}{2}\e^{a}_{\ bcd}\tilde{e}^{b-}_{\m} \star_{a}  \tilde{\omega}^{cd}_{\n}  
-  \m \leftrightarrow \n    \nn \\
\tilde{F} ^{ab}_{\m\n} &=& \p_{\m}\tilde{\omega}^{ab}_{\n} 
+  \tilde{\omega}^{a}_{\m c}\star_{s} \tilde{\omega}^{cb}_{\n} 
-4(\tilde{e}^{a+}_{\m}\star_{s}\tilde{e}^{b-}_{\n} + \tilde{e}^{a-}_{\m}\star_{s}\tilde{e}^{b+}_{\n}) 
+ i\e^{ab}_{\ \ cd}(\tilde{e}^{c+}_{\m}\star_{a}\tilde{e}^{d-}_{\n} 
-  \tilde{e}^{c-}_{\m}\star_{a}\tilde{e}^{d+}_{\n}) +  \nn \\
 &+& 2i  \tilde{a}_{\m}\star_{a}\tilde{\omega}^{ab}_{\n}  
 +  i\e^{ab}_{\ \ cd}\tilde{b}_{\m}\star_{a} \tilde{\omega}^{cd}_{\n}  -  \m \leftrightarrow \n  
 \label{curvaturenc}
\eea

Now we want to impose constraints so that the invariance is broken to $SO_{\star}(1,3)$. Moreover,
in order to recover standard gravity in the commutative limit,  the fields must satisfy 
\bea
&&\lim_{\theta\rightarrow 0}\tilde{a}_{\m} = 0 \nn \qquad\qquad
\lim_{\theta\rightarrow 0}\tilde{b}_{\m} =  0 \nn \\
&&~~~~~\nn\\
&&\lim_{\theta\rightarrow 0}\tilde{e}^{a-}_{\m} = \b ~\lim_{\theta\rightarrow 0}\tilde{e}^{a +}_{\m} 
\label{condizioniordinarie} 
\eea
To this end it is convenient to write
the gauge fields in a $\theta$-expanded form
\bea
\tilde{a}_{\m} &=&   a_{\m}^{(1)} + a_{\m}^{(2)} + ... \nn \\
\tilde{b}_{\m} &=&    b_{\m}^{(1)} +  b_{\m}^{(2)} + ... \nn \\
\tilde{e}^{a +}_{\m} &=&    e^{a +}_{\m}  +   e^{a +(1)}_{\m}  + e^{a +(2)}_{\m} +  ...\nn \\
\tilde{e}^{a -}_{\m} &=&  \b e^{a +}_{\m}  +   e^{a -(1)}_{\m}  + e^{a -(2)}_{\m} +  ...\nn \\
\tilde{\omega}^{ab}_{\m} &= & {\omega}^{ab}_{\m} +  {\omega}^{ab(1)}_{\m} +
{\omega}^{ab(2)}_{\m} + ...
\label{espansions} 
\eea
where we have taken into account the conditions in (\ref{condizioniordinarie}).  

In order to reduce the gauge symmetry we 
impose  the following constraints 
\beq
\tilde{F}^{a +} =  F^{a +} + F^{a +(1)} + ... + F^{a +(n)} +...=0
\label{torsionfree} 
\eeq
and, at each order in $\theta$, reflecting 
the different role played in the Moyal product by the even and the odd powers,
\beq
 e^{a - (n)}_{\m} =(-)^{n}\b  e^{a + (n)}_{\m} 
\label{vincoli}
\eeq
The constraints in (\ref{torsionfree}) and (\ref{vincoli}) are just sufficient to
break  $U_{\star}(2,2)$ into   $SO_{\star}(1,3)$. Indeed
 $\tilde{F}^{a +} = 0$ requires $\hat{\d}_{\tilde{\l}}\tilde{F}^{a+}_{\m\n} =0$ i.e.
\bea
\hat{\d}\tilde{F}^{a+}_{\m\n} &=&
-2\tilde{F}^{5}_{\m\n} \star_{s}\tilde{\l}^{a+}  +
4\tilde{F}^{ab}_{\m\n}\star_{s}\tilde{\l}^{+}_{ b} +  2i\tilde{F}^{1}_{\m\n}
\star_{a}\tilde{\l}^{a+}  + 2i\e^{a}_{bcd} \tilde{F}^{bc}_{\m\n}\star_{a} \tilde{\l}^{d+}\nn\\
&=&0
\eea
which leads to the condition $\tilde{\l}^{a +}=0$. With this restriction now we consider 
the constraints in (\ref{vincoli}) and impose them on the corresponding
variations $\d_{\tilde{\l}}\tilde{e}^{a+}_{\m}$ and $\d_{\tilde{\l}}\tilde{e}^{a-}_{\m}$. 
Since under a gauge transformation the connection transforms as in (\ref{transfconn})
we have
\bea
\d\tilde{e}^{a+}_{\m} &=& -2\tilde{\l}^{5} \star_{s}\tilde{e}^{a+}_{\m}  
- 4\tilde{\l}^{ab}\star_{s}\tilde{e}^{+}_{\m b} +  2i\tilde{\l}^{1} \star_{a}\tilde{e}^{a+}_{\m}  
+ 2i\e^{a}_{bcd} \tilde{\l}^{bc}\star_{a} \tilde{e}^{d+}_{\m} \nn \\
\d\tilde{e}^{a-}_{\m} &=&  \p_{\m}\tilde{\l}^{a -}  +   2 \tilde{\l}^{5} \star_{s}\tilde{e}^{a-}_{\m} 
- 4\tilde{\l}^{ab}\star_{s}\tilde{e}^{-}_{\m b} +  2i\tilde{\l}^{1} \star_{a}\tilde{e}^{a-}_{\m}  
- 2i\e^{a}_{bcd}\tilde{\l}^{bc}\star_{a} \tilde{e}^{d-}_{\m} \nn \\
&+&  2 \tilde{b}_{\m} \star_{s}\tilde{\l}^{a-} + \tilde{\omega}^{ab}_{\m}\star_{s}\tilde{\l}^{-}_{
b} +  2i\tilde{a}_{\m} \star_{a}\tilde{\l}^{a-}  - 2i\e^{a}_{bcd}\tilde{\omega}_{\m}^{bc}\star_{a}
\tilde{\l}^{d-}_{\m}  
\label{varvierbein} 
\eea
In order to satisfy (\ref{vincoli}) first we have to impose 
 $\tilde{\l}^{a-}=0$ so that the variations in (\ref{varvierbein}) become
\beq 
\d \tilde{e}^{a\pm }_{\m} =  \mp  2\tilde{\l}^{5} \star_{s}\tilde{e}^{a\pm}_{\m} -
4\tilde{\l}^{ab}\star_{s}\tilde{e}^{\pm}_{\m b}   +      2i\tilde{\l}^{1}
\star_{a}\tilde{e}^{a\pm}_{\m}  \pm     2i\e^{a}_{bcd} \tilde{\l}^{bc}\star_{a} \tilde{e}^{d
\pm}_{\m} 
\eeq 
Writing the above relations as  $\theta$-expansions
 \beq \d \tilde{e}^{a\pm
}_{\m} =  \d e^{a\pm }_{\m} +    \d e^{a\pm (1) }_{\m} + \d e^{a\pm (2)}_{\m} + ... 
\eeq
 we  obtain
\bea 
\d e^{a\pm (n) }_{\m} &=&  \mp \sum_{p+2k+q = n }   2\l^{5(p)} \star_{2k}e^{a\pm (q)
}_{\m} -  4   \sum_{p+2k+q = n } \l^{ab (p)}\star_{2k} e^{\pm (q)}_{\m b}   +\nn \\  &+&
2i \sum_{p+2k+1 +q = n }\l^{1(p)} \star_{2k+1}e^{a\pm (q)  }_{\m}  \pm     2i \sum_{p+2k+1
+q = n }  \e^{a}_{bcd} \l^{bc(p) } \star_{2k+1} e^{d \pm(q) }_{\m} 
\eea
where $p,k,q = 0,1,2,..$ and  we used the notation $ f\star g = 
\sum_{k=o}^{\infty}  f\star_{k} g$. 
At this point it is simple to show that the constraints $e^{a - (n)}_{\m} = (-)^{n}\b  e^{a + (n)}_{\m}$ 
are satisfied if 
we impose the additional conditions
\bea
\l^{1(2n)} &=& \l^{5(2n)} = 0 \nn \\
\l^{ab (2n+1)} &=& 0 
\eea
Therefore the  restricted gauge parameter $\tilde{\l}$ belongs to $SO_{\star}(1,3)$
and this completes our proof.

The action  \cite{Nair:2001kr} which is invariant under an $SO_{\star}(1,3)$ transformation
$\tilde{\l} = \tilde{\l}^{1}i\1 + \tilde{\l}^{5}\gamma_{5}+ \tilde{\l}^{ab}\gamma_{ab}$ is
\beq 
S_{NC} = \int d^{4}x ~\e^{\m\n\r\s} Tr(\gamma_{5}\tilde{F}_{\m\n}\star \tilde{F}_{\r\s})
\label{azioneNC} 
\eeq
Indeed one immediately obtains
 \beq
\hat{\d}_{\tilde{\l}}S_{NC} = \int d^{4}x ~\e^{\m\n\r\s} Tr([\gamma_{5} ,
\tilde{\l}]\star  \tilde{F}_{\m\n}\star\tilde{F}_{\r\s}) = 0 
\eeq
More explicitly (\ref{azioneNC}) can be rewritten as
 \beq
S_{NC} =   \frac{i}{2} \int d^{4}x~\e^{\m\n\r\s} \  \bigg{(} 16 \tilde{F}^{1}_{\m\n}
\tilde{F}^{5}_{\r\s} - \e_{abcd}\tilde{F}^{ab}_{\m\n} \tilde{F}^{cd}_{\r\s} \bigg{)} 
\label{azioneNC2}
\eeq
with field strengths as given in (\ref{curvaturenc}).


Now we want to use the  constraints (\ref{torsionfree}) and (\ref{vincoli}) in (\ref{azioneNC2}) and express the dependent fields in terms of the independent, dynamical ones. First we use
\beq \tilde{F}^{a+}_{\m\n}  = \p_{\m}\tilde{e}^{a +}_{\n}  - 2\tilde{b}_{\m}
\star_{s}\tilde{e}^{a+}_{\n}  + \tilde{\omega}^{ab}_{\m} \star_{s}\tilde{e}^{+}_{\n b}  +
2i\tilde{a}_{\m} \star_{a}\tilde{e}^{a+}_{\n}  + \frac{i}{2}\e^{a}_{bcd}\tilde{e}^{b+}_{\m} \star_{a}  
\tilde{\omega}^{cd}_{\n}  -  \m \leftrightarrow \n  = 0
\nn 
\eeq
and  determine $\tilde{\omega}^{ab}_{\m}$ order by order in $\theta$,
thus obtaining  $\omega^{ab(n)}_{\m} = \omega^{ab(n)}_{\m}(e_{\m}^{a+},...,e_{\m}^{a+(n)}$,
 $a_{\m}^{(1)},...,a_{\m}^{(n -1)}$, $b_{\m}^{(1)},...,b_{\m}^{(n)})$.
 
At this level we have obtained a theory in terms of the fields 
 $\tilde{e}_{\m}^{a+}$, $\tilde{a}_{\m}$ and $\tilde{b}_{\m}$
 invariant under transformations $\tilde{\l}\in SO_{\star}(1,3)$. It represents a noncommutative deformation of Einstein gravity which contains the vierbeins plus an infinite number of additional fields which enter at all orders in the $\theta$-expansion of $\tilde{e}_{\m}^{a+}$, $\tilde{a}_{\m}$ and $\tilde{b}_{\m}$. Now we attempt to reduce the number of independent fields 
 employing the Seiberg-Witten map \cite{SW}.\\
In general the map allows to express the gauge connection of a noncommutative theory $\hat{A}_{\m}$ as a  a $\t$-expansion of standard  gauge theory variables $A_{\m}$. In the present case we want to identify the fields  $\tilde{e}_{\m}^{a+}$, $\tilde{a}_{\m}$ and $\tilde{b}_{\m}$ with the corresponding ones obtained via the Seiberg-Wittten maps $\hat{e}_{\m}^{a+}(e_{\m}^{a+}) $, $\hat{a}_{\m}(e_{\m}^{a+})  $ and $\hat{b}_{\m}(e_{\m}^{a+})$.\\
 As we will show the procedure is consistent with the choice of the constraints (\ref{torsionfree}) and (\ref{vincoli}).
 The only dynamical fields of the theory turn out to be the vierbeins $e_{\m}^{a+}$ in terms of which we can define  the space-time metric \beq 
g_{\m\n}= e_{\m}^{a+} e_{\n a}^{+}
\eeq
In order to implement consistently this procedure it is crucial to prove that
 the gauge group $SO_{\star}(1,3)$
of the noncommutative theory is related to the $SO(1,3)$ commutative theory precisely via the 
Seiberg-Witten map. Then we can determine
the functions $\tilde{e}_{\m}^{a+}(e_{\m}^{a+})$, $\tilde{a}_{\m}(e_{\m}^{a+})$ and
$\tilde{b}_{\m}(e_{\m}^{a+})$ solving the equations \cite{SW}
 \bea
\hat{\d}_{\hat{\l}}\tilde{e}_{\m}^{a+}(e_{\m}^{a+}) &=& \tilde{e}_{\m}^{a+}(e_{\m}^{a+} 
+ \d_{\l}e_{\m}^{a+}) - \tilde{e}_{\m}^{a+}(e_{\m}^{a+}) \nn \\
\hat{\d}_{\hat{\l}}\tilde{a}_{\m}(e_{\m}^{a+})&=& \tilde{a}_{\m}(e_{\m}^{a+} +\d_{\l}e_{\m}^{a+}) 
- \tilde{a}_{\m}(e_{\m}^{a+} ) \nn \\
\hat{\d}_{\hat{\l}}\tilde{b}_{\m}(e_{\m}^{a+})&=& \tilde{b}_{\m}(e_{\m}^{a+}
+\d_{\l}e_{\m}^{a+}) - \tilde{b}_{\m}(e_{\m}^{a+} ) 
\eea
where $\l$ belongs to the Lorentz algebra, i.e. $\l = \l^{ab}\gamma_{ab}$ (the $SO(1,3)$ 
gauge group of the commutative limit), while $\hat{\l}$ belongs to the $SO_{\star}(1,3)$
gauge group of the corresponding (via the Seiberg-Witten map) noncommutative theory.

Thus let us show that if we start from an SO(1,3) gauge theory  and use the Seiberg-Witten map
to construct the corresponding noncommutative one, the gauge group 
is precisely mapped into $SO_{\star}(1,3)$. We describe the commutative theory through its
connection  $A_{\m} = \frac{1}{4}\omega^{ab}_{\m}\gamma_{ab}$ and gauge
parameters $\l = \l^{ab}\gamma_{ab}$ and the corresponding noncommutative one through
$\hat{A}(A)$ and $\hat{\l}(\l,A)$ defined as \cite{SW}
 \bea
\d_{\hat{\l}}\hat{A}_{\m} =& \hat{A}_{\m}(A + \d_{\l}A) - \hat{A}_{\m}(A) 
\label{sw1}
\eea
where
 \bea
\d_{\l}A_{\m} &=&  \p_{\m}  \l   +   \l A_{\m} - A_{\m} \l \nn \\
\d_{\hat{\l}}\hat{A}_{\m} &=&   \p_{\m}  \hat{\l}   +   \hat{\l}\star \hat{A}_{\m} 
- \hat{A}_{\m}\star \hat{\l} 
\label{sw2}
\eea 
The solution of (\ref{sw1}) is equivalent to 
 \bea
\d \hat{A}_{\m}(\t) &=& -\frac{i}{4} \d \t^{\a\b} \{ \hat{A}_{\a}, ( \p_{\b}\hat{A}_{\m}
 + \hat{F}_{\b\m}) \}_{\star} \nn \\
\d\hat{\l}(\t) &=& \frac{i}{4} \d \t^{\a\b} \{ \p_{\a} \l, A_{\b} \}_{\star}.
\label{sw3} 
\eea
Our goal is to prove that the gauge parameter $\hat{\l}$ belongs to $SO_{\star}(1,3)$. 
We look for a solution of (\ref{sw3}) in a $\theta$-expanded form:
\bea
\hat{A}_{\m} &=& {A}_{\m} + {A}_{\m}^{(1)} +{A}_{\m}^{(2)} +...\nn \\
\hat{\l} &=& \l + \l^{(1)} + \l^{(2)} +... 
\label{thetaex}
\eea
and we want to prove that
 \bea
 {A}_{\m}^{(n)}(A) &=&  {A}_{\m}^{(n)I}(A)t_{I(n)} \nn\\
\lambda^{(n)}(A) &=& \lambda^{(n)I}(A)t_{I(n)} 
 \label{str}
\eea
where $t_{I(n)} = \gamma_{ab}$ if $n$ is even, while  $t_{I(n)} \in (i\1, \gamma_{5})$ if $n$ is odd.
We  prove (\ref{str}) by induction first for ${A}_{\m}^{(n)}$.
We begin with  the $n=1$ case.
As emphasized  above in the commutative theory we have
$A_{\m} = \frac{1}{4}\omega^{ab}_{\m}\gamma_{ab}$, $\l = \l^{ab}\gamma_{ab}$ and 
$ F_{\m\n} = \frac{1}{4}F^{ab}_{\m\n}\gamma_{ab}$.
Thus inserting $\d{A}_{\m}^{(1)} =
{A}_{\m \a\b} \d \t^{\a\b}$  in the first equation   (\ref{sw3}) we obtain the result for $n=1$
\beq 
{A}_{\m \a\b}  =  -\frac{1}{2}  A_{\a}^{ab} ( \p_{\b}A_{\m}^{cd} + F_{\b\m}^{cd})
(\eta_{ac}\eta_{bd}i\1 + \e_{abcd}\gamma_{5})
\eeq
Now assuming the result is true for general $n$, we prove it for $n+1$.
We define
 \beq 
 {A}_{\m}^{(n)} = \frac{1}{n}{A}_{\m
\a_{1}\b_{1},...,\a_{n}\b_{n}}\t^{\a_{1}\b_{1}}...\t^{\a_{n}\b_{n}} 
\eeq 
so that
\beq
\d{A}_{\m}^{(n)}(\t) = {A}_{\m \a_{1}\b_{1},...,\a_{n}\b_{n}}\t^{\a_{1}\b_{1}}...\d
\t^{\a_{n}\b_{n}}
 \eeq
Inserting the above expressions  in the first equation (\ref{sw3}) we find 
\bea
\d{A}_{\m}^{(n+1)}(\t) 
&=&    -\frac{i}{4} \d \t^{\a\b}\sum_{p+2k+q=n}  A_{\a}^{(p)I} \star_{2k}  ( \p_{\b}A_{\m}^{(q)J} 
+ F_{\b\m}^{(q)J})  \{t_{I(p)}, t_{J(q)} \} + \nn \\
 &-& \frac{i}{4} \d \t^{\a\b}\sum_{p+2k+1+q=n}  A_{\a}^{(p)I}\star_{2k+1} ( \p_{\b}A_{\m}^{(q)J} 
 + F_{\b\m}^{(q)J}) [t_{I(p)}, t_{J(q)}] \label{relazione}
\eea
Using the standard commutation relations among the generators we end up with
${A}_{\m}^{(2n)} = {A}_{\m}^{(2n)ab}\gamma_{ab}$ and 
${A}_{\m}^{(2n+1)} = {A}_{\m}^{(2n+1)1}i\1 + {A}_{\m}^{(2n+1)5}\gamma_{5}$.
This result on the $\t$-structure of $\hat{A}_{\m}(\t)$, and  the second of the
Seiberg and Witten equations (\ref{sw3}) lead to the conclusion that
the parameter $\hat{\l}(\t)$ belongs to $SO_{\star}(1,3)$.

As anticipated above now we can safely determine the independent fields of our noncommutative theory
through the Seiberg-Witten map.
We apply this procedure and explicitly compute the first order noncommutative correction
to the standard gravity action.
The action is expanded in powers of $\theta$
\bea
S_{NC} & =&   \frac{i}{2} \int d^{4}x~\e^{\m\n\r\s} \  \bigg{(} 
16 \tilde{F}^{1}_{\m\n} \tilde{F}^{5}_{\r\s} - \e_{abcd}
\tilde{F}^{ab}_{\m\n} \tilde{F}^{cd}_{\r\s} \bigg{)}  \nn  \\
&=&   S + S^{(1)} +  S^{(2)} + ... 
\eea 
and the first noncommutative correction $S^{(1)}$ is evaluated in terms of the 
dynamical fields $e_{\m}^{a+}$ as follows:
from (\ref{condizioniordinarie}) and (\ref{curvaturenc}) we find 
that  $F^{1}_{\m\n} = F^{5}_{\m\n} = 0$. 
Thus $S^{(1)}$ is simply given by
\beq
 S^{(1)} = -i \int d^{4}x~\e^{\m\n\r\s} \e_{abcd} F^{ab(1)}_{\m\n} F^{cd}_{\r\s}
  \label{s1}
\eeq 
Inserting the constraints
\bea
&&a_{\m} = b_{\m} =0 \nn \\
&&e_{\m}^{a-} = \b e_{\m}^{a+} \qquad\qquad
e_{\m}^{a-(1)} = -\b e_{\m}^{a+(1)} 
\eea
in the expression (\ref{curvaturenc}) for $\tilde{F}^{ab}_{\m\n}$, we obtain 
\bea
F^{ab}_{\m\n} &=&  \p_{\m}\omega^{ab}_{\n} +  \omega^{a}_{\m c} \omega^{cb}_{\n} - 
8 \b e^{a+}_{\m}e^{b+}_{\n} -  \m \leftrightarrow \n \nn \\
F^{ab(1)}_{\m\n} &=& \p_{\m}\omega^{ab(1)}_{\n} +  \omega^{a(1)}_{\m c} \omega^{cb}_{\n} +
\omega^{a}_{\m c} \omega^{cb(1)}_{\n} - \m \leftrightarrow \n \label{F^1}
\eea
Now solving  the constraints $ F_{\m\n}^{a+}  = 0$  in  $\omega^{ab}_{\m}$ 
 and $ F_{\m\n}^{a+(1)} = 0$  in  $\omega^{ab(1)}_{\m}$ we determine the spin connections 
 in terms of the vierbein  $e_{\m}^{a+}$.

At $0^{th}$-order in $\theta$ we have
 \beq
  F_{\m\n}^{a+}  =  \p_{\m}e_{\n}^{a+} - \p_{\n}e_{\m}^{a+}  + \omega^{ab}_{\m}{e}^{+}_{\n b} -   
  \omega^{ab}_{\n}{e}^{+}_{\m b}   = 0
\eeq
 It is solved by
\beq
 \omega^{ab}_{\m} = -\frac{1}{2}\e^{\r a +} \e^{\n b +}( \p_{\m}e_{\n}^{k+}e_{\r k}^{+} - 
 \p_{\n}e_{\r}^{k+}e_{\m k}^{+} +  \p_{\r}e_{\m}^{k+}e_{\n k}^{+}) 
 \label{omega}
\eeq 
where  $\e^{\m +}_{a}$ is the inverse vierbein
 \beq
e^{a+}_{\n}\e^{\n +}_{b} = \d^{a}_{b} \qquad\qquad
e^{a+}_{\m}\e^{\n +}_{a} = \d^{\n}_{\m}
 \eeq
 
In the same way  at $1^{st}$-order we have
 \beq 
 F^{a+(1)}_{\m\n}  =
\p_{\m}e^{a +(1)}_{\n}  - 2b_{\m}^{(1)} e^{a+}_{\n}  + \omega^{ab(1)}_{\m} e^{+}_{\n b} +
\omega^{ab}_{\m} e^{+(1)}_{\n b}   -  \frac{1}{4}\t^{\a\b}\e^{a}_{\ bcd}\p_{\a} e^{b+}_{\m}
\p_{\b} \omega^{cd}_{\n}  -  \m \leftrightarrow \n  = 0 
\label{1order}
\eeq 
With the definition
 \beq
   A^{a(1)}_{\m \n} \equiv      \p_{\m}e^{a +(1)}_{\n}  - 2b_{\m}^{(1)} e^{a+}_{\n}  
   +  \omega^{ab}_{\m} e^{+(1)}_{\n b}   -  \frac{1}{4}\t^{\a\b}\e^{a}_{\ bcd}\p_{\a} e^{b+}_{\m} \p_{\b} 
   \omega^{cd}_{\n}
\eeq 
we  rewrite (\ref{1order}) as
 \beq
 \omega^{ab(1)}_{\m} e^{+}_{\n b} -  \omega^{ab(1)}_{\n} e^{+}_{\m b}  = - A^{a(1)}_{\m \n} 
 + A^{a(1)}_{\n \m}
\eeq 
Therefore we obtain
 \beq 
 \omega^{ab(1)}_{\m} = -\frac{1}{2} \e^{\r a +} \e^{\n b +}  
(A^{k(1)}_{\m \n} e_{\r k}^{+} -  A^{k(1)}_{\n \r} e_{\m k}^{+} + A^{k(1)}_{\r \m} e_{\n
k}^{+})
\label{omega1} 
\eeq
At this stage we have $\omega^{ab}_{\m}(e^{a+}_{\m})$ from (\ref{omega}) and  
$\omega_{\m}^{ab(1)}(e^{a+}_{\m},e^{a+(1)}_{\m},b_{\m}^{(1)})$ from (\ref{omega1}).

We still have to evaluate $b_{\m}^{(1)}$ and  $e^{a+(1)}_{\m}$ in terms of
$ e_{\m}^{a+}$.      
We do this via the  Seiberg-Witten map: in addition to the relations in (\ref{sw2}) and 
(\ref{sw3}) we have correspondingly
 \bea
\d_{\l}e_{\m}^{+} &=&    \l e_{\m}^{+} - e_{\m}^{+} \l \nn \\
\d_{\hat{\l}}\hat{e}^{+}_{\m} &=&     \hat{\l}\star \hat{e}_{\m}^{+} -
\hat{e}_{\m}^{+}\star \hat{\l}
\label{swe1} 
\eea 
and \cite{Bonora:2000td}
\beq
\d \hat{e}_{\m}^{+}(\t) = -\frac{1}{2}\d \t^{\a\b} ( \{ \hat{A}_{\a},  \p_{\b}\hat{e}^{+}_{\m}\}_{\star}
+ \frac{1}{2}\{ [\hat{e}^{+}_{\m},  \hat{A}_{\a}]_{\star},  \hat{A}_{\b} \}_{\star} ) 
\label{swe2} 
\eeq
From (\ref{sw3}) we have to $1^{st}$-order
\beq
  A_{\m}^{(1)} = -\frac{i}{4}  \t^{\a\b} \{ A_{\a}, ( \p_{\b}A_{\m} + F_{\b\m}) \}
\eeq 
Substituting in the above equation 
 $A_{\m} = \frac{1}{4}\omega^{ab}_{\m}\gamma_{ab}$,  $\l = \l^{ab}\gamma_{ab}$, $F_{\m\n} 
 = \frac{1}{4}F^{ab}_{\m\n}\gamma_{ab}$, with $\omega^{ab}_{\m}$ as given in (\ref{omega})
we obtain
 \bea
A_{\m}^{(1)} &=& -\frac{i}{16}  \t^{\a\b}\omega^{ab}_{\a}( \p_{\b}\omega^{cd}_{\m} 
+ F^{cd}_{\b\m}) \{ \gamma_{ab}, \gamma_{cd} \} \nn \\
&=&  -\frac{i}{16}  \t^{\a\b}\omega^{ab}_{\a}( \p_{\b}\omega^{cd}_{\m} + F^{cd}_{\b\m})
( -2i\eta_{ac}\eta_{bd}i\1 -2i \e_{abcd}\gamma_{5}) \nn \\
&=& a_{\m}^{(1)}i\1 +  b_{\m}^{(1)}\gamma_{5} 
\eea 
In this way
\bea
 a_{\m}^{(1)} &=& -\frac{1}{8}  \t^{\a\b}\omega^{ab}_{\a}( \p_{\b}\omega_{ \m ab} + F_{\b\m ab }) \nn \\
b_{\m}^{(1)}& =&
 -\frac{1}{8}  \t^{\a\b} \e_{abcd}  \omega^{ab}_{\a}(
\p_{\b}\omega^{cd}_{\m} + F^{cd}_{\b\m}) 
\label{a(1) e b(1)}
 \eea
are determined as functions of $e_{\m}^{a+}$.

In the same way from (\ref{swe2}) we obtain
\bea
e^{+(1)}_{\m}(e,A) &=& -\frac{i}{2}\t^{\a\b}\bigg{(}\{ A_{\a}, \p_{\b} e^{+}_{\m} \}
+\frac{1}{2}\{ [  e^{+}_{\m}, A_{\a}],  A_{\b}\} \bigg{)}\nn\\ 
&=&  -\frac{i}{2}\t^{\a\b}\bigg{(}\frac{1}{4}
\omega_{\a}^{bc}\p_{\b}e^{a+}_{\m}  \{i\gamma_{a}^{+}, \gamma_{bc} \} +\frac{1}{2}\{
e^{a+}_{\m} \frac{1}{4} \omega_{\a}^{bc}   [ i\gamma_{a}^{+},\gamma_{bc} ] , \frac{1}{4}
\omega_{\b}^{ef}\gamma_{ef} \}\bigg{)} 
\eea 
Using 
\bea
\lbrack  i\gamma_{a}^{+} , \gamma_{bc} \rbrack &=&   2\eta_{ab}i\gamma_{c}^{+} 
- 2\eta_{ac}i\gamma_{b}^{+}  \nn \\
\{ i\gamma_{a}^{+} ,  \gamma_{bc} \} &=&2i\e_{abcd}(i \gamma_{d}^{+} ) 
\eea
 finally we obtain: 
\beq
  e^{a+(1)}_{\m} = \frac{1}{4}\t^{\a\b}\e^{a}_{bcd} ( \p_{\b}e^{b+}_{\m} \omega_{\a}^{cd} 
  - \frac{1}{4}\omega_{\a k}^{b}e^{k+}_{\m} \omega_{\b}^{cd}). \label{e(1)}
\eeq
Using the expressions given in (\ref{e(1)}) and in (\ref{a(1) e b(1)}) we can reconstruct the first order correction of the spin-connection  (\ref{omega1}) which finally allows to obtain  $F^{ab(1)}_{\m\n}$ in (\ref{F^1}).
In this way one can  reexpress the first order correction in (\ref{s1}) explicitly in terms of the vierbeins $e^{a+}_{\m}$.\\
The complete result requires a straightforward but quite lengthy algebra. It would be interesting to proceed further and  write the action explicitly in terms of the metric. Then one could evaluate the corrected propagator and
 investigate how these $\theta$-dependent terms  affect the renormalization properties of the theory.

\vspace{0.8cm}


\noindent {\bf Acknowledgements}

\noindent We thank S. Cacciatori, L. Martucci and M. Picariello for very helpful comments and suggestions.\\ 
 This work has been partially supported by INFN, MURST, and the European Commission RTN program 
HPRN-CT-2000-00113 in which the authors are associated to the University of Torino.

\end{document}